\documentclass[review]{elsarticle}

\usepackage{lineno,hyperref}
\modulolinenumbers[5]

\usepackage{amssymb}
\usepackage{epsfig}
\usepackage{graphicx}
\usepackage{times}
\usepackage{float}
\usepackage[usenames,dvipsnames]{color}
\usepackage{multirow,amsmath}
\usepackage{booktabs}

\newcommand{\ALPHALogo}{\includegraphics[height=32pt]{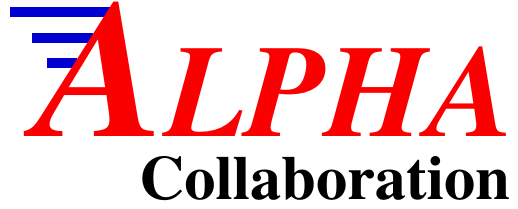}}

\newcommand{\eq}[1]{eq.~(\ref{#1})}

\newcommand{\fig}[1]{Fig.~\ref{#1}}

\newcommand{\sect}[1]{Sect.~\ref{#1}}

\newcommand{\tab}[1]{Table~\ref{#1}}

\newcommand{\MeV}{{\rm MeV}}

\newcommand{\fm}{{\rm fm}}

\newcommand{\ZP}{Z_{\rm P}}

\newcommand{\mbarsf}{\kern1pt\overline{\kern-1pt m\kern-1pt}\kern1pt_{{\rm SF}}}
\newcommand{\gbarsf}{\kern1pt\overline{\kern-1pt g\kern-1pt}\kern1pt_{{\rm SF}}}

\newcommand{\Lameff}{\Lambda_\mathrm{dec}}

\newcommand{\msbar}{{\rm \overline{MS\kern-0.05em}\kern0.05em}}
\newcommand{\lag}[1]{{\mathcal{L}}_{\rm {#1}}}
\newcommand{\rmO}{{\rm O}}
\newcommand{\mhad}{m^\mathrm{had}}

\newcommand{\mbar}{\kern1pt\overline{\kern-1pt m\kern-1pt}\kern1pt}
\newcommand{\Mc}{M_{\rm c}}

\newcommand{\rmd}{{\rm d}}
\def\fm{{\rm fm}}

\newcommand{\ev}[1]{\left\langle #1 \right\rangle}
\newcommand{\rmA}{{\rm A}}
\newcommand{\nf}{N_\mathrm{f}}

\newcommand{\LamYM}{\Lambda_{\rm YM}}
\newcommand{\mhadYM}{\mhad_{\rm YM}}

\begin{document}

\begin{frontmatter}

\title{
\begin{flushright}
\small{
WUB/17-02\\
HU-EP-17/15\\
DAMTP-2017-22}
\vskip 0.7cm
\end{flushright}
Power corrections from decoupling of the charm quark}

\author{\ALPHALogo \hfill\\Francesco Knechtli, Tomasz Korzec}
\address{Fakult\"at f\"ur Mathematik und Naturwissenschaften, Bergische Universit\"at Wuppertal,\\ Gau{\ss}str. 20, 42119 Wuppertal, Germany}

\author{Bj{\"o}rn Leder}
\address{Institut f\"ur Physik, Humboldt Universit\"at,\\ 
Newtonstr. 15, 12489 Berlin, Germany}

\author{Graham Moir}
\address{DAMTP, Centre for Mathematical Sciences,\\
Wilberforce Road, CB3 0WA Cambridge, UK}

\begin{abstract}
Decoupling of heavy quarks at low energies can be described by means 
of an effective theory as shown by S.~Weinberg in Ref.~[1]. We 
study the decoupling of the charm quark by lattice simulations. We 
simulate a model, QCD with two degenerate charm quarks. In this case the
leading order term in the effective theory is a pure gauge theory. The 
higher order terms are proportional to inverse powers of the charm quark 
mass $M$ starting at $M^{-2}$. Ratios of hadronic scales are equal to their 
value in the pure gauge theory up to power corrections. We show, by 
precise measurements of ratios of scales defined from the Wilson flow,
that these corrections are very small and that they can be described by a term 
proportional to $M^{-2}$ down to masses in the region of the charm quark mass.
\end{abstract}


\end{frontmatter}


\section{Introduction}
\label{sec:intro}

In a field theory which contains light (mass-less) fields and fields of a
heavy mass $M$, the functional integral over the latter can be performed
resulting in an effective theory for the light fields which was formulated by 
Weinberg \cite{Weinberg:1980wa}. The action of the effective theory contains
the action of the light fields (without the heavy fields) and an infinite
number of non-renormalizable terms. The latter are suppressed by powers of
$E/M$ at low energies $E\ll M$. Moreover, the non-renormalizable couplings do 
not contribute to the renormalization group equations of the renormalizable
couplings of the light fields. This property holds for mass-independent 
renormalization schemes like the $\msbar$ scheme as shown in 
\cite{Weinberg:1980wa}. The heavy fields still
affect the value of the renormalized couplings of the light fields through 
the decoupling relations, which result from the matching of the effective 
and the fundamental theory at low energies.

Assuming the validity of perturbation theory at the matching scale,
the decoupling relations can be computed perturbatively.
In the case of QCD and one heavy quark, such as the charm or the bottom quark,
the decoupling relation for the renormalized strong coupling is known to
four loops \cite{Weinberg:1980wa,Bernreuther:1981sg,Chetyrkin:2005ia,Schroder:2005hy}.
The strong coupling of the five-flavor theory can be extracted in this way 
from the coupling computed non-perturbatively in the three-flavor theory 
using lattice simulations \cite{Bruno:2017gxd}.
We remark that
the decoupling relation for the strong coupling can be equivalently expressed 
as a relation between the $\Lambda$ parameters of the effective and the
fundamental theory \cite{Bruno:2014ufa}.

Simulations of QCD on the lattice are often carried out with three light 
sea quarks
\cite{Lin:2008pr,Aoki:2009ix,Bietenholz:2010jr,Arthur:2012yc,Bruno:2014jqa,Bali:2016umi}.
The inclusion of a charm sea quark increases significantly the computational
cost and introduces additional tuning to set the bare quark masses on a line
of constant physics. Moreover, in the case of simulations with Wilson 
fermions, Symanzik O($a$) improvement requires the computation of coefficients 
which multiplies terms proportional to the bare quark masses in lattice units 
$am$ \cite{Luscher:1996sc,Bhattacharya:2005rb}. 
The contribution of these terms is significant
for the charm quark $am_c>0.3$ at the affordable lattice spacings $a>0.05\fm$.
Some of these coefficients, like the one of the gluon action, are difficult
to extract non-perturbatively.
Relying on decoupling of the charm quark at low energies allows to simulate
the cheaper and simpler effective theory with three flavors only. 

The applicability of decoupling for the charm quark has to be justified. 
In \cite{Bruno:2014ufa} this was studied in a model, QCD with two 
heavy mass-degenerate quarks and no light quarks.
The decoupling of the heavy quarks leave a pure gauge theory\footnote{
Perturbatively the simultaneous decoupling of two heavy quarks is known 
at three-loop order \cite{Grozin:2011nk}.}
up to power corrections (which are due to the non-renormalizable interactions)
at low energies. The latter were extracted by computing
low energy quantities related to the Wilson
flow \cite{Atiyah:1982fa,Narayanan:2006rf,Luscher:2009eq,Lohmayer:2011si}.
Ratios of two such quantities are insensitive to the matching of
the gauge couplings, and after taking the continuum limit, can be 
compared to their counterparts in the pure gauge theory. The differences
are due to the power corrections.
By interpolating data obtained from simulations at quark
masses ranging from $1/8$th up to one half of the charm quark mass with data
from simulations of the pure gauge theory, the size of the power corrections 
due to one sea charm quark was estimated to be at the sub-percent level
\cite{Bruno:2014ufa}.

In \cite{Bruno:2014ufa} it was noted that the simulated masses were not
large enough to see the leading behavior of the power corrections
which start at $1/M^2$ in the effective theory. 
Instead a behavior more like $1/M$ was observed.
In this article we study the same model as in \cite{Bruno:2014ufa} but extend 
the simulated quark masses to the charm quark mass and slightly above. 
Thus we can directly compute the size of the power corrections from decoupling 
of the charm quark. Furthermore
we perform a non-perturbative test of the validity of the
effective theory of decoupling for the charm quark.
Our goal is to determine whether the leading power corrections in the 
inverse heavy quark mass behave in the charm region as $1/M^2$.

The article is organized as follows. In \sect{sec:dec} we briefly
review the theoretical framework of the 
effective theory of decoupling for QCD with two heavy mass-degenerate quarks (in the continuum). 
\sect{sec:ens} presents the details of the Monte Carlo simulations of this
model formulated on the lattice.
The results for the ratios of low energy quantities are presented
in \sect{sec:res} and their dependence on the heavy quark mass is compared to
the effective theory prediction. The conclusions of our work are drawn in
\sect{sec:concl}.

\section{Decoupling}
\label{sec:dec}

To avoid a multi-scale problem, we consider a simplified version of QCD, namely 
an $SU(3)$ Yang--Mills theory coupled to two degenerate heavy quarks. 
This allows us to perform simulations in relatively small volumes with very 
small lattice spacings, as we describe in \sect{sec:ens}. 
We briefly review the theoretical framework of decoupling specifically
for our model. The fundamental theory is QCD with $\nf=2$ mass-degenerate
quarks. $\Lambda$ is the Lambda parameter in the $\msbar$ scheme
and $M$ is the renormalization group invariant (RGI) mass of the heavy
quarks.\footnote{
Throughout this work, the $\Lambda$ parameter is
defined in the $\msbar$ scheme. For mass-independent schemes like the $\msbar$, 
there is an exact one-loop relation for the $\Lambda$ parameters between
different schemes. The RGI mass $M$ is independent of the scheme (for 
mass-independent schemes).}
After decoupling of the heavy quarks, what is left is a pure gauge theory.
Therefore, the Lagrangian of the effective theory valid at energies $E\ll M$ 
is given by \cite{Weinberg:1980wa,chir:Weinberg}
\begin{equation}\label{e:Leff}
\lag{dec} = \lag{YM}
+ 1/M^{2} \sum_i \omega_i \Phi_i +\rmO(\Lambda^{4}/M^{4}) \,.
\end{equation}
$\lag{YM}$ is the Lagrangian of the $SU(3)$ Yang--Mills (pure gauge) theory.
Due to gauge invariance there are no fields of mass dimension equal to five.
A complete set of fields of mass dimension equal to six is
$\Phi_1=\mathrm{tr}\,\{D_\mu F_{\nu\rho} D_\mu F_{\nu\rho}\}$ and
$\Phi_2=\mathrm{tr}\,\{D_\mu F_{\mu\rho} D_\nu F_{\nu\rho}\}$, where
$F_{\mu\nu}$ is the $SU(3)$ field strength tensor and $D_\mu F_{\nu\rho}$ its
covariant derivative.

At leading order the effective theory, \eq{e:Leff}, is a Yang--Mills theory.
It has only one free parameter, the renormalized gauge coupling.
This coupling is fixed by matching the effective theory to the fundamental
theory. Equivalently one can fix the $\Lambda$ parameter of the Yang--Mills 
theory, $\LamYM$, which becomes a function 
$\LamYM=\Lameff(M,\Lambda)$, see \cite{Bruno:2014ufa,dec:longpaper}.
Matching requires that low energy physical observables
are the same in the two theories up to power corrections. 
Let us denote a low energy observable by $\mhad$ where, for example,
it can represent a hadronic scale such as $1/\sqrt{t_0}$ \cite{Luscher:2010iy}
or $1/r_0$ \cite{Sommer:1993ce}.
After matching
\begin{equation}\label{e:matching}
\mhad(M) = \mhadYM + \rmO(\Lambda^{2}/M^{2})~,
\end{equation}
where $\mhad(M)$ is the hadronic scale in QCD with $\nf=2$ heavy quarks
of mass $M$ and $\mhadYM$ is the hadronic scale in the Yang--Mills theory.
Note that $\mhadYM$ depends on $M$ through the
matching, in particular $\mhadYM/\LamYM$
is a pure number independent of $M$. Therefore
we consider two hadronic scales,
$m^{\mathrm{had,1}}(M)$ and $m^{\mathrm{had,2}}(M)$, whose values in the
Yang--Mills theory are $m_{\rm YM}^{\mathrm{had,1}}$ and
$m_{\rm YM}^{\mathrm{had,2}}$ respectively. An immediate consequence of
\eq{e:matching} is
\begin{equation}\label{e:ratio}
R(M) = \frac{m^{\mathrm{had,1}}(M)}{m^{\mathrm{had,2}}(M)}
     = \frac{m_{\rm YM}^{\mathrm{had,1}}}{m_{\rm YM}^{\mathrm{had,2}}} + 
       \rmO(\Lambda^{2}/M^{2}) \,.
\end{equation}
The matching of the coupling
is irrelevant for the ratios and we have direct access to the
power corrections \cite{Knechtli:2014sta}.
The effective theory of decoupling predicts that the ratios like in 
\eq{e:ratio} are equal to their value $R(M=\infty)$ in the Yang--Mills theory
with a leading power correction in the inverse heavy quark mass given by
\begin{equation}\label{e:powercorr}
R(M) = R(\infty) + k \Lambda^{2}/M^{2} \,,
\end{equation}
where $k$ is a parameter which depends on the hadronic scales which are
taken to form the ratio.
The goal of this work is to verify the behavior in \eq{e:powercorr} and
to establish whether it applies for masses around the charm quark mass.

\section{Monte Carlo simulations}
\label{sec:ens}

\begin{table}[t]
\centering
{\small
\begin{tabular}{c c c c c c c c}
\toprule
$\frac{T}{a}\times\left(\frac{L}{a}\right)^3$ &  $\beta$  & $\kappa$    & $a \mu$            & $M/\Lambda$ & $r_0/a$   & $t_0/a^2$ & MDUs \\
\midrule
$120\times 32^3$                              &  5.300    & 0.136457    & 0.024505           & 0.5900      &   --      & 4.174(13) & 4300 \\
$120\times 32^3$                              &  5.500    & 0.1367749   & 0.018334           & 0.5900      & 8.77(15)  & 7.917(82) & 8000 \\
$192\times 48^3$                              &  5.700    & 0.136687    & 0.013713           & 0.5900      &   --      & 14.40(10) & 5800 \\
\midrule
$120\times 32^3$                              &  5.500    & 0.1367749   & 0.039776           & 1.2800      & 8.010(62) & 6.871(33) & 8000 \\
$192\times 48^3$                              &  5.700    & 0.136687    & 0.029751           & 1.2800      &   --      & 12.668(39)& 16200\\ 
\midrule
$120\times 32^3$                              &  5.500    & 0.1367749   & 0.077687           & 2.5000      & 7.392(62) & 5.836(27) & 8000 \\
$192\times 48^3$                              &  5.700    & 0.136687    & 0.058108           & 2.5000      &   --      & 10.916(38)& 9000 \\
\midrule
$192\times 48^3$                              &  5.600    & 0.136710    & 0.130949           & 4.8700      &   --      & 6.609(15) & 2000 \\
$120\times 32^3$                              &  5.700    & 0.136698    & 0.113200           & 4.8703      & 9.123(57) & 9.104(36) & 17184 \\
$192\times48^3$                               &  5.880    & 0.136509    & 0.087626           & 4.8700      & 11.946(55) & 15.622(62)& 23088 \\
$192\times 48^3$                              &  6.000    & 0.136335    & 0.072557           & 4.8700      & 14.34(10) &22.39(12)& 22400 \\
\midrule
$192\times 48^3$                              &  5.600    & 0.136710    & 0.155367           & 5.7781      &   --      & 6.181(11)& 2096 \\
$192\times 48^3$                              &  5.700    & 0.136687    & 0.1343             & 5.7781      &   --      & 8.565(31) &  2700 \\
$120\times 32^3$                              &  5.880    &  0.136509   & 0.103965           & 5.7781      &   --      &14.916(93)  &  59888\\
\midrule
$120\times 32^3$                              &  6.100    &    --       &    --              & $\infty$    & 6.345(13) & 4.4329(32)& 64000 \\
$120\times 32^3$                              &  6.340    &    --       &    --              & $\infty$    & 9.029(77) & 9.034(29) & 20080 \\
$120\times 24^3$                              &  6.340    &    --       &    --              & $\infty$    &   --      & 9.002(31) & 60920 \\
$192\times 48^3$                              &  6.672    &    --       &    --              & $\infty$    & 14.103(92) & 21.924(81)& 73920 \\
$192\times 64^3$                              &  6.900    &    --       &    --              & $\infty$    & 18.93(15) &39.41(15)&160200 \\
\bottomrule
\end{tabular}
}
\caption{Simulation parameters of our twisted mass and quenched ensembles. 
The columns show the lattice sizes,
the gauge coupling $\beta=6/{g_0^2}$, 
the hopping parameter (for maximal twist), 
the twisted mass parameter,
the ratio of the RGI mass to the $\Lambda$ parameter ($\infty$ for quenched),
the scales $r_0/a$ (where it is measured) and $t_0/a^2$ and 
the total statistics in molecular dynamics units. 
}\label{t:ens}
\end{table}

We simulate QCD with two mass-degenerate flavors of quarks ($\nf=2$).
Wilson's plaquette gauge action~\cite{Wilson:1974sk} is employed in the
the Yang-Mills sector and a doublet of quarks is
realized either as standard or as twisted mass~\cite{Frezzotti:2000nk} Wilson quarks.
In both cases a clover term~\cite{Sheikholeslami:1985ij,Luscher:1996sc} with 
non-perturbatively determined improvement coefficient $c_{\rm sw}$~\cite{Jansen:1998mx} is
added. It is not needed for the $O(a)$ improvement of the twisted mass action
at maximal twist, but was found to reduce the $O(a^2)$ lattice artifacts,
see e.g.~\cite{Dimopoulos:2009es}.

The bare coupling $\beta$ of the gauge action was chosen such that the 
lattice spacings 
cover the range $0.023~{\rm fm}\lesssim a \lesssim 0.066~{\rm fm}$.
The lattice spacing is determined from the hadronic scale $L_1$ 
\cite{Blossier:2012qu,Fritzsch:2012wq}.
The scale $L_1/a$ is defined at vanishing quark mass, where
the standard and twisted mass Wilson quark formulations are
equivalent. Therefore, the lattice spacing for a given bare coupling $\beta$ is
the same for both formulations.
In order to obtain the scale $L_1$ in 
lattice units at a particular value of $\beta$, we fitted the data
in Table 13 of~\cite{Fritzsch:2012wq} as it is explained there.
The lattice spacing in physical units is estimated by rescaling
the value $a=0.0486\,\mathrm{fm}$ at $\beta=5.5$ from \cite{Fritzsch:2012wq}
by the ratio of the $L_1/a$ values.

In order to resolve the short correlation lengths associated with the large quark
masses that we aim at, we are forced to simulate at very small lattice spacings.
Critical slowing down becomes a major obstacle which we alleviate by the implementation of
open boundary conditions in the time direction~\cite{Luscher:2011kk}. The 
boundary improvement coefficients are kept at their 
tree-level values  $c_G=1$ and $c_F=1$.
The publicly available
{\tt openQCD} simulation program~\cite{openQCD,algo:openQCD} is used for
our simulations.

We used standard O($a$) improved Wilson quarks to simulate at quark masses of
approximately a factor $1/8$, $1/4$ and $1/2$ of the charm quark mass. 
The details of these simulations are given
in \cite{Bruno:2014ufa,dec:longpaper}.
For the present work we also simulated twisted mass Wilson quarks
at these quark masses. In addition, we also simulated directly at the charm
quark mass and at one mass larger than that of the charm quark.
In the simulations of twisted mass Wilson quarks,
the hopping parameter $\kappa$ was set to its critical value in order to 
achieve maximal twist. The critical
values were obtained from an interpolation of published data~\cite{Fritzsch:2012wq,Blossier:2012qu,Fritzsch:2015eka}.
The twisted mass parameter $a\mu$ was
chosen to correspond to certain values of $M/\Lambda$ listed in \tab{t:ens}.
More precisely, at a 
given value of the bare coupling the twisted mass parameter is set by
\begin{equation}
   a\mu = \frac{M}{\Lambda}\, \ZP(L_1)\, \frac{\bar m(L_1)}{M} \, \Lambda L_1 \, \frac{a}{L_1}\, .
   \label{Eq:amu}
\end{equation}
The pseudo-scalar renormalization constant at renormalization 
scale $L_1^{-1}$ computed in the Schr\"odinger 
Functional scheme $\ZP(L_1)=0.5184(53)$ (valid for $5.2\le\beta\le6.0$), 
the relation between the running and the RGI mass 
$M/\bar m(L_1) = 1.308(16)$ and $\Lambda = 310(20)\,\MeV$
are known from~\cite{Fritzsch:2012wq,DellaMorte:2005kg}. 
We take $\Lambda L_1=0.649(45)$ from~\cite{DellaMorte:2004bc}.
Some of the quantities entering eq.~(\ref{Eq:amu}) have rather large errors.
The dominant error comes from $\Lambda L_1$. Note that it is common to all our
simulation points and amounts to a change in the target values of $M/\Lambda$.
For the charm quark we set $\Mc/\Lambda=4.8700$, where we use
the preliminary value $\Mc=1510\,\MeV$ of \cite{Heitger:2013oaa}
which agrees with \cite{Olive:2016xmw}.

In order to determine the power corrections in \eq{e:ratio}, we also simulate
the pure gauge theory at values of the scales $t_0/a^2$ \cite{Luscher:2010iy},
$r_0/a$ \cite{Sommer:1993ce} which are similar or larger. 

Table~\ref{t:ens} summarizes our twisted mass and quenched ensembles. 
At the large quark masses of our simulations, the system is close to 
a pure gauge theory, with similar finite volume effects.
Our lattice volumes are such that $L/\sqrt{t_0}\ge8$. We explicitly checked
in the pure gauge theory that a volume $L/\sqrt{t_0}=8$ is large enough
to exclude finite volume effects in the scales derived from the Wilson
flow ($t_0$, $t_c$, $w_0$, see \sect{sec:res}). In fact, we can rule out 
significant finite volume effects in all data which we use in the analysis in 
\sect{sec:res}.

\section{Results}
\label{sec:res}

On the ensembles generated with two flavors of
O($a$) improved Wilson fermions reported in \cite{Bruno:2014ufa,dec:longpaper} 
and those generated in the twisted mass simulations 
at maximal twist and the pure gauge simulations which are listed in \tab{t:ens},
 we measure
the hadronic scales $\sqrt{t_0}$, $\sqrt{t_c}$ and $w_0$. They are defined from
gauge fields smoothed by the Wilson flow 
\cite{Narayanan:2006rf,Luscher:2009eq,Lohmayer:2011si} as follows.
We denote by $E(x,t)$ the smoothed action density, where $t$ is the flow time
of mass dimension $-2$, and introduce the dimensionless quantity
\begin{equation}\label{e:density}
{\cal E}(t) = t^2 \ev{E(x,t)} \,.
\end{equation}
At flow times $t>0$, $E(x,t)$ is a renormalized quantity 
\cite{Luscher:2010iy,Luscher:2011bx}.
The reference scale $t_0$ is defined as in Ref.~\cite{Luscher:2010iy} by
\begin{equation}\label{e:t0}
{\cal E}(t_0) = 0.3 \,.
\end{equation}
Similarly, the scale $t_c$ is defined by
\begin{equation}\label{e:tc}
{\cal E}(t_c) = 0.2 \,.
\end{equation}
The numerical solutions to \eq{e:t0} and \eq{e:tc} are found
by quadratic interpolation of the data of ${\cal E}(t)$.
We use the clover (symmetric) definition of the action density $E$ on
the lattice, cf. \cite{Luscher:2010iy}.
The scale $w_0$ is defined as in Ref.~\cite{Borsanyi:2012zs} by
\begin{equation}\label{e:w0}
w_0^2 {\cal E}^\prime(w_0^2) = 0.3 \,,
\end{equation}
where ${\cal E}^\prime(t) = \frac{\rmd}{\rmd t}{\cal E}(t)$.
The numerical solution to \eq{e:w0}
is found by first computing the symmetric finite differences of 
$t^2E(t)$ on each configuration and then by quadratic interpolation of 
the data.
The error analysis is based on the method of Ref.~\cite{Wolff:2003sm}
and takes into account the coupling to slow modes following Ref.~\cite{Schaefer:2010hu}.
We estimate the exponential autocorrelation time $\tau_{\rm exp}$
from the tail of the autocorrelation function of $t_0$.
We use the ensembles at our largest masses, for which we perform simulations at our
finest lattice spacings. As expected with open boundary conditions~\cite{Luscher:2011kk},
the observed critical slowing down is compatible with a $\tau_{\rm exp}\propto a^{-2}$
behavior and from a least squares fit we obtain
$\tau_{\rm exp} = -32(23) + 17.4(2.8)\,t_0/a^2$ in molecular dynamics units 
(MDUs). The errors of the fit coefficients are given in parantheses.
With periodic boundary conditions the autocorrelation times would be
much larger, cf.~\cite{Schaefer:2010hu} where an effective scaling 
proportional to $a^{-5}$ was observed.

We compute the ratios of hadronic scales
\begin{eqnarray}
 \label{e:ratios}
R = \sqrt{t_c/t_0} \; & \mbox{and} \; & R = \sqrt{t_0}/w_0 \,.
\end{eqnarray}
In such ratios the bare coupling (or equivalently the lattice spacing) drops
out. After taking the continuum limit we can directly compare the ratios
in the $\nf=2$ theory to their value in the pure gauge theory and so determine 
the size of the $1/M^{2}$ effects in \eq{e:powercorr}.
\begin{figure}[t]\centering
  \includegraphics[height=7cm]{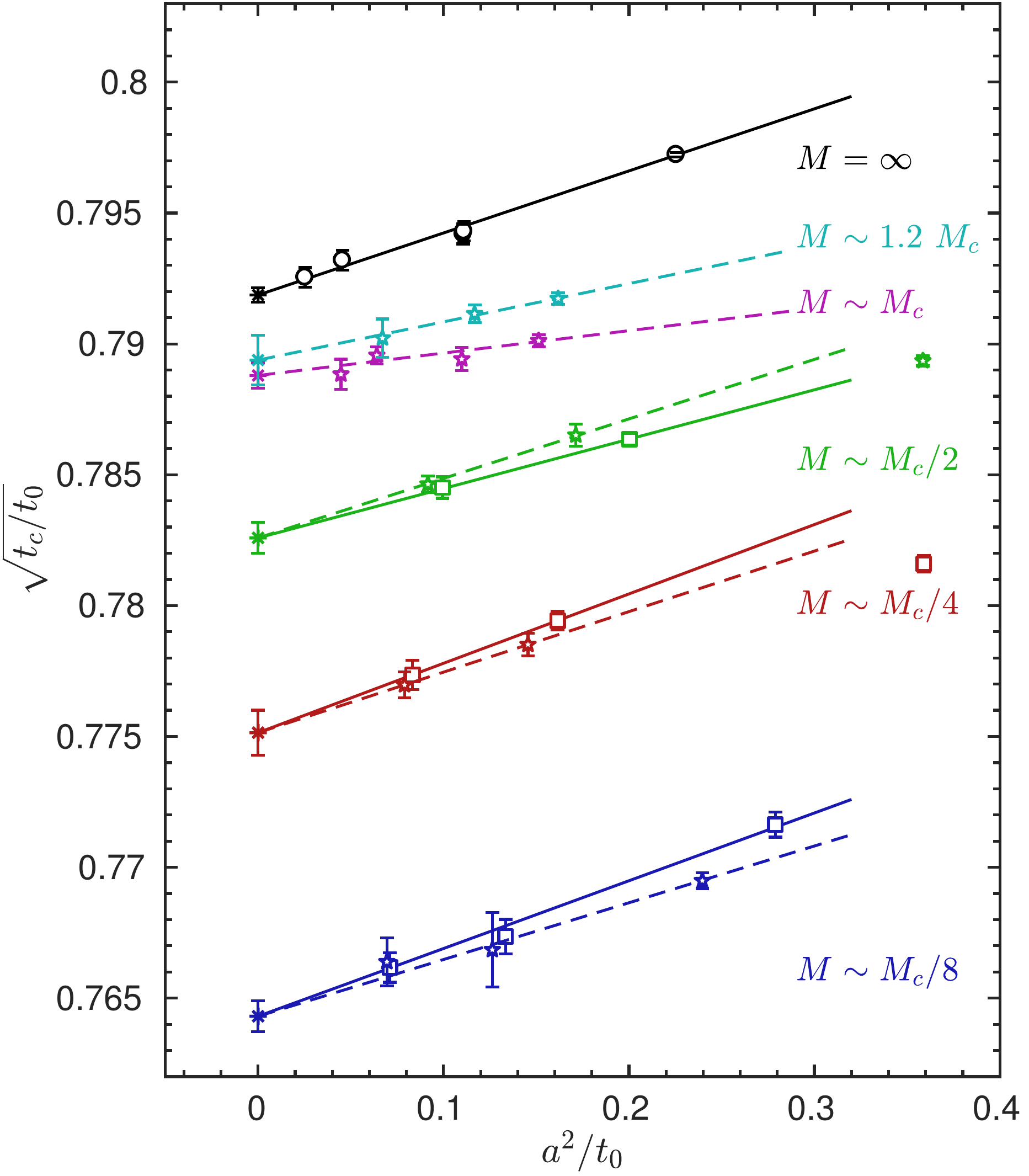}\hfill 
  \includegraphics[height=7cm]{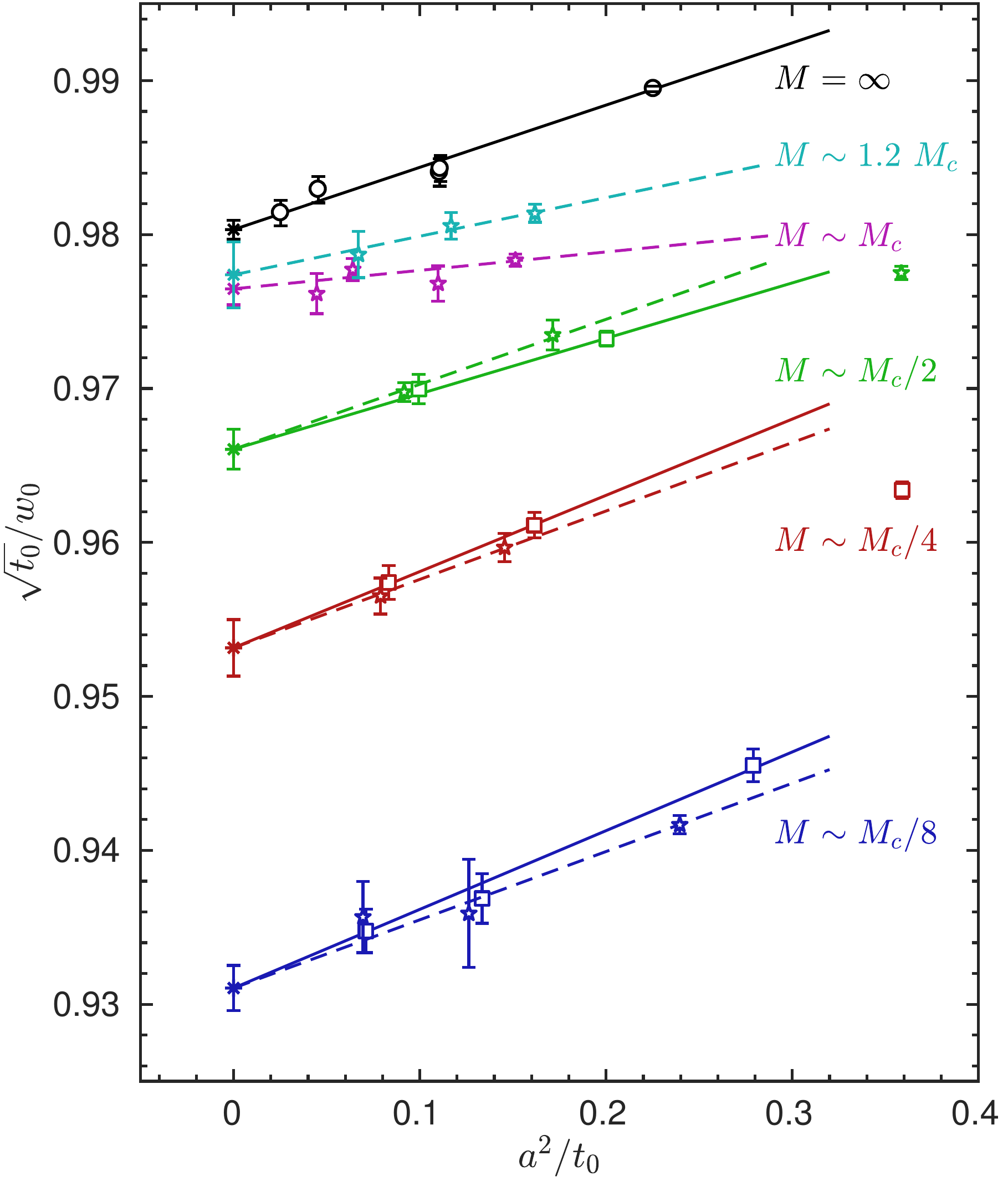}
  \caption{Combined continuum extrapolations. 
  On the left, the ratio $\sqrt{t_c/t_0}$. On the right $\sqrt{t_0}/w_0$.
We show data from twisted mass (pentagrams), standard Wilson (squares) and
quenched (circles) simulations.
For the second coarsest quenched lattice we performed a finite volume test
and there are two data points overlapping.
The lines represent the continuum extrapolations described in
the text and the asterisks are the obtained continuum values.}
  \label{f:tct0}
\end{figure}
\begin{figure}[t]\centering
  \includegraphics[height=7cm]{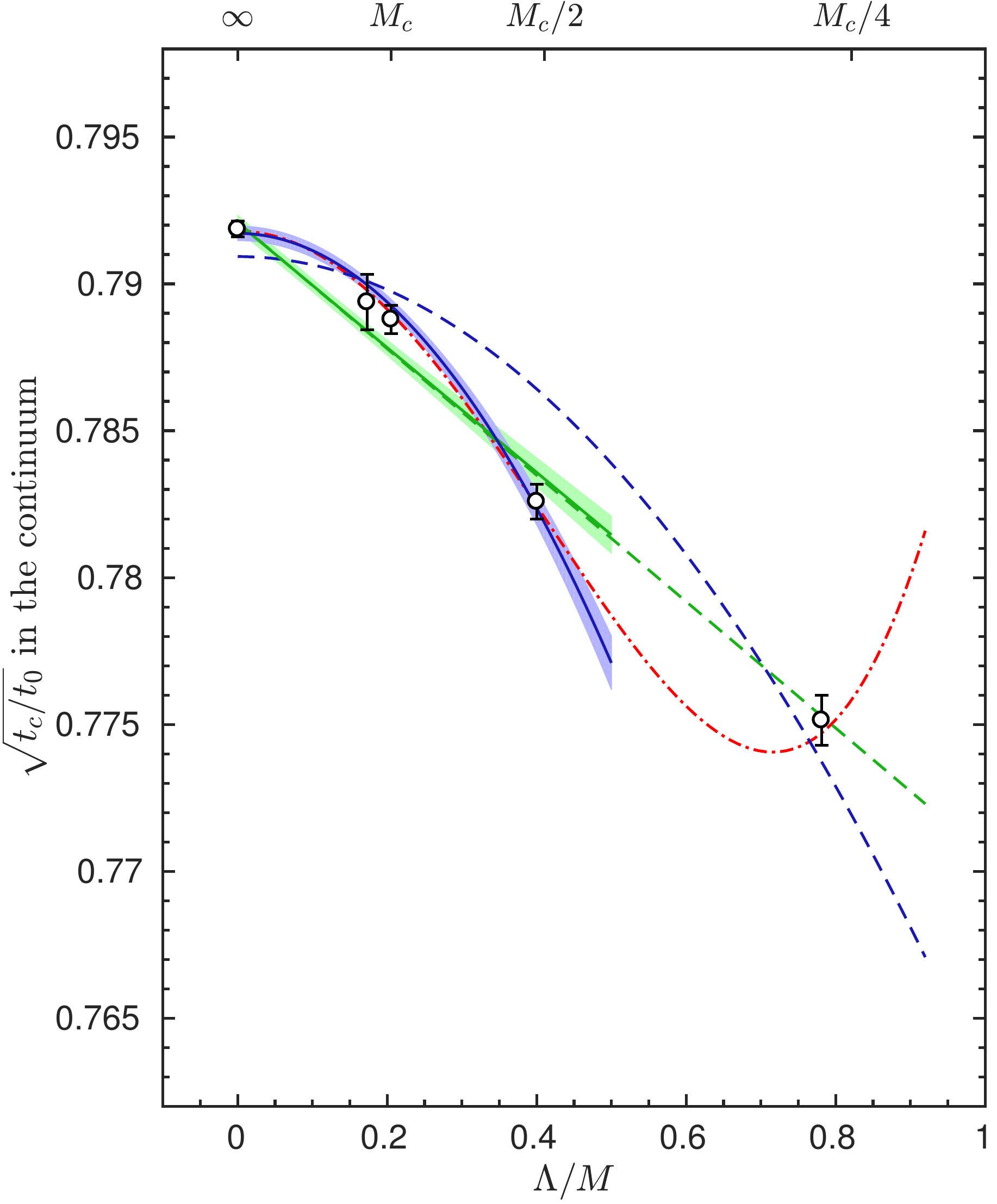}\hfill 
  \includegraphics[height=7cm]{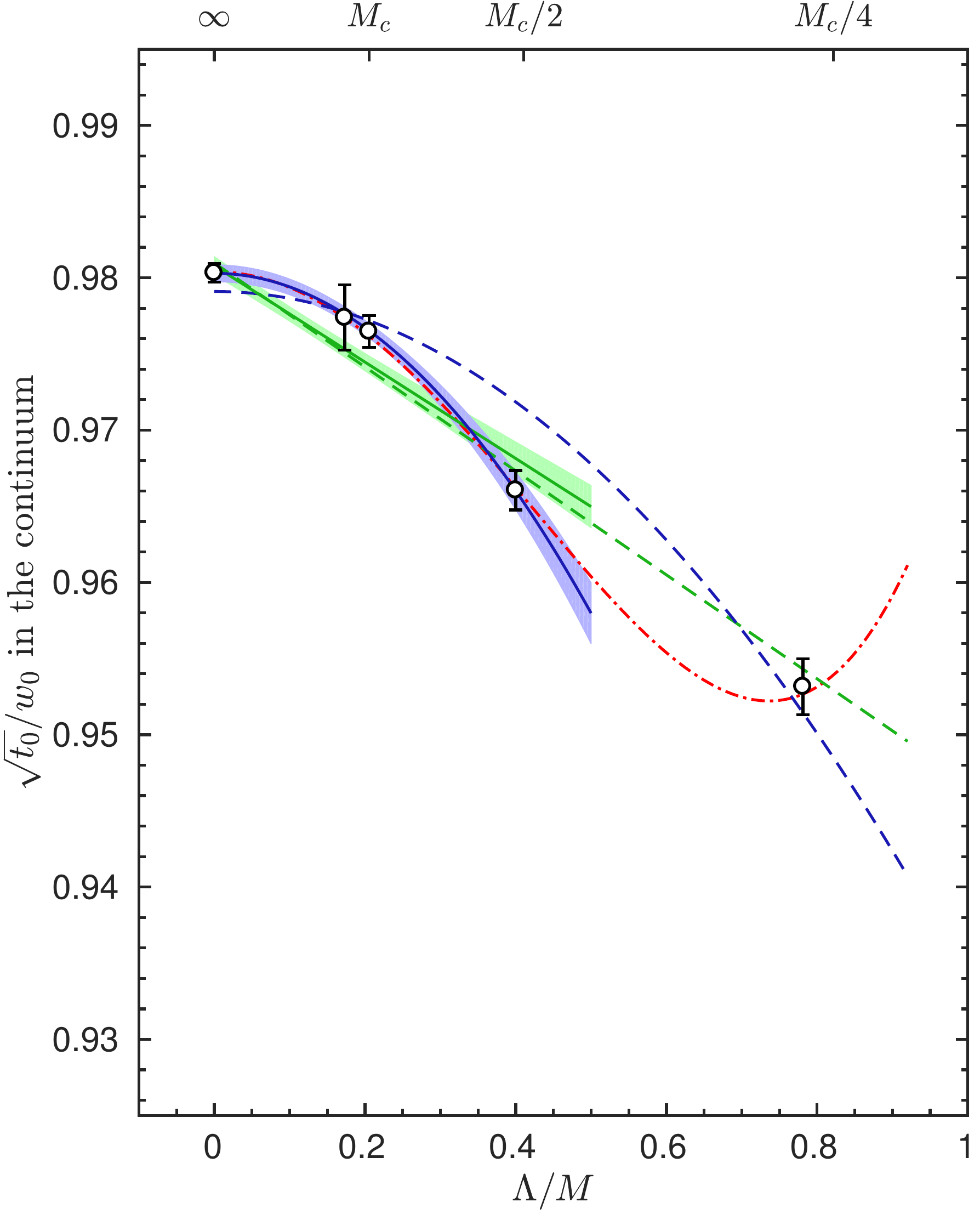}
  \caption{The continuum extrapolated values of $\sqrt{t_c/t_0}$ (left) and $\sqrt{t_0}/w_0$ (right) from the
fit shown in \fig{f:tct0} plotted against $\Lambda/M$. 
The line in the blue band is the effective theory prediction 
\eq{e:powercorr} fitted 
through points from $M=\infty$ down to $M/\Lambda=2.5000$.
The line in the green band is instead a fit linear in $\Lambda/M$.
For comparison the dashed lines represent the quadratic (blue) and linear
(green) fit through points from $M=\infty$ down to $M/\Lambda=1.2800$.
Also shown by the dashed-dotted red line is a fit in this range
adding to \eq{e:powercorr} a next-to-leading correction term proportional 
to $\Lambda^4/M^4$.}
  \label{f:tct0_cont}
\end{figure}

To determine the values of the ratios in the continuum limit, we fit our data to
$R(a,M/\Lambda,\rmA) = R^{\rm cont}(M/\Lambda) + \frac{a^2}{t_0} c(M/\Lambda, \rmA)$,
where A is the action, ``W'' for Wilson, ``tm'' for twisted mass and ``q'' for quenched (pure gauge, $M=\infty$).
The functional form is motivated by Symanzik's effective theory for our actions. For a given mass
$M/\Lambda$ we have two fit parameters (continuum value and slope) for cases where the
calculation was performed with one action and three parameters (continuum value and two slopes) 
for cases with two lattice actions.
We apply a cut, $a^2/t_0(M)<0.32$, to the data to be fitted.
The data and the fits for $R=\sqrt{t_c/t_0}$ and $R=\sqrt{t_0}/w_0$ are shown in 
\fig{f:tct0}. 
The pentagrams represent the twisted mass data, the squares are the 
standard Wilson data
and the circles are the pure gauge data. 
The lines represent fits: the dashed lines are the continuum extrapolations
of the twisted mass data and the continued lines are the continuum 
extrapolations of the Wilson and of the quenched data.
The asterisks represent the
continuum extrapolated values $R^{\rm cont}(M/\Lambda)$.
The data is described very well by the fits and the continuum values are very 
stable under changes in the fitting procedure, such as removing some of the data points,
or changes in the cut in $a^2/t_0(M)$. Moreover, a global fit that models the $M$ dependence of
the slopes yields compatible values. Although the global fit has less parameters, we prefer 
the individual fits since they yield statistically independent continuum values.

\begin{table}[H]
\centering
{\small
\begin{tabular}{c c c c c c c}
\toprule
$M/\Lambda$      & $\infty$  & 5.7781     & 4.87      & 2.50      & 1.28        & 0.59 \\
\midrule
$\sqrt{t_c/t_0}$ & 0.7919(3) & 0.7894(9)  & 0.7888(5) & 0.7826(6) & 0.7751(9)  & 0.7643(6)\\ 
$\sqrt{t_0}/w_0$ & 0.9803(6) & 0.9774(21) & 0.9765(10)& 0.9661(13)& 0.9532(18)  & 0.9311(15)\\
$r_0/\sqrt{t_0}$ & 3.013(17) & -          & 3.022(29) & 2.988(35) & 3.043(71)   & 3.050(64)\\
\bottomrule
\end{tabular}
}
\caption{The values of various dimensionless ratios in the continuum limit for several values
of the quark mass.}\label{t:Rcont}
\end{table}

The continuum values $R^{\rm cont}(M/\Lambda)$ obtained from the fit are
summarized in \tab{t:Rcont} and are plotted
in \fig{f:tct0_cont} against $\Lambda/M$.
The relative effect of a single charm quark compared to pure gauge theory
can be estimated by 
$1/2[R^{\rm cont}(\Mc)-R^{\rm cont}(\infty)]/R^{\rm cont}(\infty)$. 
This effect is very small, it is $-0.00196(37)$ for $R = \sqrt{t_c/t_0}$ and
$-0.00194(59)$ for $R = \sqrt{t_0}/w_0$.
In \fig{f:tct0_cont} the line in the blue
band represents the effective theory prediction \eq{e:powercorr}, where the
coefficient $k$ of the $(\Lambda/M)^2$ term is determined by a fit that includes
data down to $M/\Lambda = 2.50$.
Our data are very well described by \eq{e:powercorr}.
A fit to a behavior of the power corrections linear in $M^{-1}$ is shown
by the line in the green band. 
A linear behavior in $1/M$ was observed in the data for masses smaller than
half of the charm quark mass~\cite{Bruno:2014ufa,dec:longpaper}.
While adding the new data for larger masses cannot exclude the $M^{-1}$ 
behavior completely, this fit is far worse than the one with $M^{-2}$ 
corrections.
The fitted values of the coefficients $k$ and the $\chi^2$ values per degree of 
freedom of the fits
are listed in \tab{t:k} . 
The second error of $k$ is systematic and is given by the difference between 
fits where the lowest mass included in the fit is $\Mc$ or $\Mc/2$.

\begin{table}[H]
\centering
\begin{tabular}{@{\extracolsep{0.2cm}}cccccc}
\toprule
$R$   & \multicolumn{3}{c}{fits down to $\Mc/2$} &
\multicolumn{2}{c}{fits down to $\Mc/4$} \\
\cmidrule(lr){2-4}\cmidrule(lr){5-6} 
& $k$ & $\chi^2_{\rm quad} / {\rm dof}$ & $\chi^2_{\rm lin} / {\rm dof}$ 
& $\chi^2_{\rm quad} / {\rm dof}$ & $\chi^2_{\rm lin} / {\rm dof}$ \\
\midrule
$\sqrt{t_c/t_0}$ & -0.058(04)(16) & 1.75 / 2 & 9.55 / 2 &  65.76 / 3 & 9.61 / 3
\\
$\sqrt{t_0}/w_0$ & -0.089(09)(03) & 0.02 / 2 & 8.54 / 2 &  27.31 / 3 & 9.54 / 3
\\
$r_0/\sqrt{t_0}$ & -0.14(24)(37)  & 0.25 / 1 & 0.42 / 1 &   0.78 / 2 & 0.79 / 2
\\ \bottomrule
\end{tabular}
\caption{The $\chi^2$ values of the quadratic \eq{e:powercorr}
and linear fits in $\Lambda/M$ for different ratios
shown in Figure~\ref{f:tct0_cont} and on the right plot in Figure~\ref{f:r0t0}.
We perform two types of fits, the first through the $M=\infty$ and 
$M_c/2$ data points and the second extending down to the $M_c/4$ data points.
For the first type of fits we list the values of the coefficients $k$ of the 
$(\Lambda/M)^2$ term in \eq{e:powercorr}.
The first error is statistical and the second is systematic.
}\label{t:k}
\end{table}

In \fig{f:tct0_cont} we also show the fits to the data down to masses 
corresponding to $\Mc/4$. They are represented by the dashed lines,
blue for the quadratic and green for the linear fit.
While the linear fit is almost unchanged with respect to the linear fit 
down to $\Mc/2$, the quadratic fit is clearly excluded. Still the quadratic 
fit to the points down to $\Mc/2$ has much better $\chi^2$ values per degree 
of freedom than the linear fit down to $\Mc/4$, see 
\tab{t:k} . These results demonstrate that the linear behavior is disfavored 
only when the data for masses smaller than $\Mc/2$ are excluded from the fits. 
The data then begin to show the leading quadratic behavior which is expected 
from the effective theory for large enough masses. This conclusion is
further supported by the dashed-dotted lines shown in \fig{f:tct0_cont}. 
They represent a fit down to $\Mc/4$ when a next-to-leading correction
term $\Lambda^4/M^4$ (whose coefficient is fitted) is added to the leading 
behavior of \eq{e:powercorr}. This fit has very good $\chi^2$ values per 
degree of freedom: $0.49 / 2$ for $R = \sqrt{t_c/t_0}$ and
$0.11 / 2$ for $R = \sqrt{t_0}/w_0$. It deviates substantially from the
fit to the leading behavior down to $\Mc/2$ only for masses smaller 
than $\Mc/2$.
\begin{figure}[t]\centering
  \includegraphics[height=7cm]{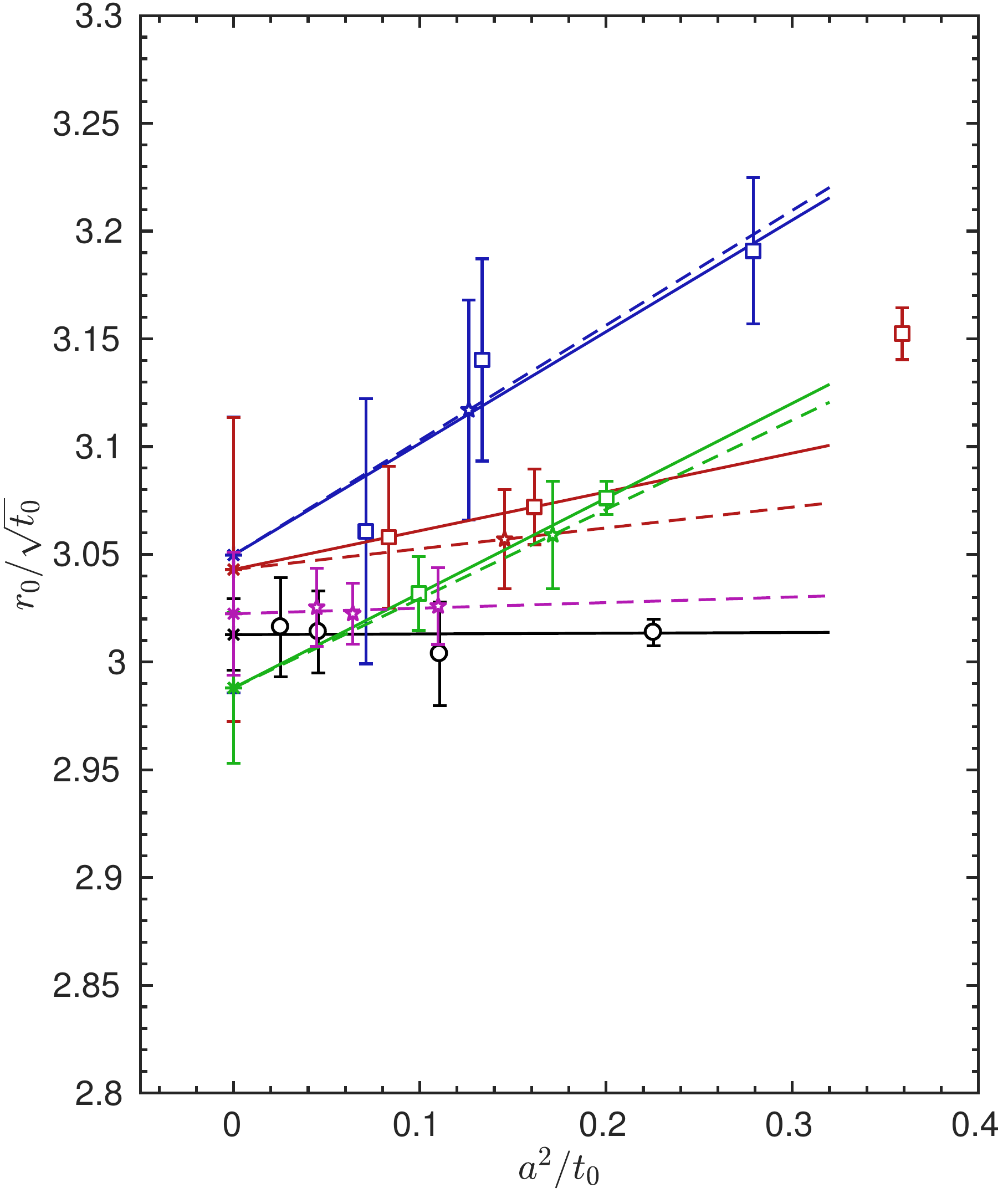}\hfill 
  \includegraphics[height=7cm]{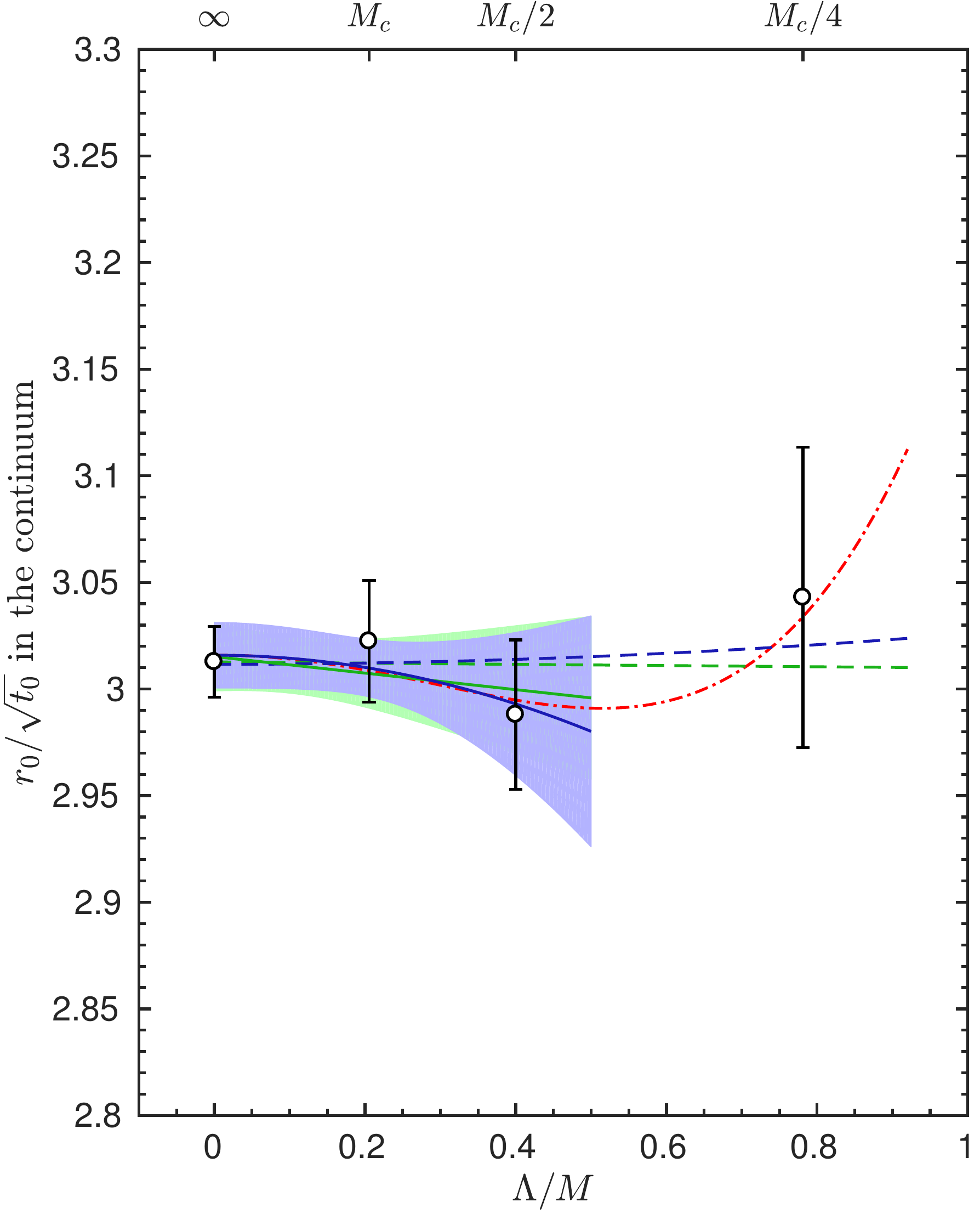}
  \caption{Combined continuum extrapolations of the ratio $r_0/\sqrt{t_0}$ (left) and the mass
           dependence of the continuum values (right). The symbols and colors are chosen as in 
           \fig{f:tct0} and \fig{f:tct0_cont}.}
  \label{f:r0t0}
\end{figure}

In order to appreciate the precision that can be reached with flow quantities, we also present
the corresponding results for the ratio $R=r_0/\sqrt{t_0}$ including the 
Sommer-scale $r_0$ \cite{Sommer:1993ce} extracted from Wilson loops. 
For the measurements of Wilson loops we use the {\tt wloop} package\footnote{
It is available at {\tt https://github.com/bjoern-leder/wloop/}.}
which implements the method of~\cite{Donnellan:2010mx}. This amounts to firstly smearing all gauge
links (for the temporal links this means a choice of the static action) and subsequently measuring the
Wilson loops where the initial and final line of gauge links 
are smeared using up to four levels of spatial
HYP smearing \cite{Hasenfratz:2001hp}. This allows us to extract the static-quark potential $a\,V(r)$ very reliably by solving a generalized 
eigenvalue problem~\cite{Blossier:2009kd}. 
The static force $F(r) = V'(r)$ can then be used to measure the hadronic scale $r_0$ defined implicitly 
through \cite{Sommer:1993ce}
\begin{equation}\label{e:r0}
r_0^2 F(r_0) = 1.65 \,.
\end{equation}
Even a careful state of the
art determination of $r_0$ does not yield a precision high enough to resolve the power corrections
we are interested in, as can be seen in \fig{f:r0t0}.
We note that the coefficient $k$ of the $(\Lambda/M)^2$ term 
in \eq{e:powercorr} depends on the observables used to form the ratio.

\section{Conclusions}
\label{sec:concl}

In this work
we simulated a model, QCD with two heavy mass-degenerate quarks.
At low energies this theory is described by an effective theory 
which is a pure gauge theory up to power corrections in the inverse 
heavy quark mass.
By comparing ratios of low energy physical quantities computed in both
theories and extrapolated to the continuum limit, we could determine the
size of the power corrections. 
They have been found to be very small. We are now confident that the effects 
of neglecting the charm quark in $\nf=2+1$ simulations is far below a percent 
in dimensionless low-energy quantities.
The power corrections are
expected for sufficiently large heavy quark masses to be proportional to the 
square of the inverse quark mass, see \eq{e:powercorr}.
Our data shown in \fig{f:tct0_cont} follow very well this expectation
down to masses equal to half of the charm quark mass.
A behavior linear in the inverse quark mass, which is possible for masses
outside the range of validity of the effective theory, is strongly disfavored
by the data.

In order to achieve a stronger conclusion and find the range of quark masses 
where the linear behavior can be completely excluded, the statistics of our 
simulations should be increased and larger quark masses up to the bottom quark 
mass be simulated. The resources needed to carry this out are beyond our 
computational budget. We emphasize that the computational resources used to 
produce the data for this article amount to a large scale project already. 
Simulating with a yet larger statistical precision and heavier quark masses 
would require a computational effort which is comparable to simulations with
light sea quarks and are therefore beyond the scope of this model calculation.

\vspace{0.5cm}

\noindent
{\bf Acknowledgements.}
The study of decoupling was initiated by Rainer Sommer whom we thank
for many useful discussions and for his valuable comments on the manuscript. 
We also thank Jacob Finkenrath for a check on our measurements.
The authors gratefully acknowledge the Gauss Centre for Supercomputing (GCS) 
for providing computing time for a GCS Large-Scale Project on the GCS share of 
the supercomputer JUQUEEN at J\"ulich Supercomputing Centre (JSC). GCS is the 
alliance of the three national supercomputing centres HLRS (Universit\"at 
Stuttgart), JSC (Forschungszentrum J\"ulich), and LRZ (Bayerische Akademie der 
Wissenschaften), funded by the German Federal Ministry of Education and 
Research (BMBF) and the German State Ministries for Research of 
Baden-W\"urttemberg (MWK), Bayern (StMWFK) and Nordrhein-Westfalen (MIWF). GM
acknowledges support from the Herchel Smith Fund at the University of Cambridge.
This work is supported by the Deutsche Forschungsgemeinschaft in the SFB/TR 55.





\bibliography{charm}

\end{document}